\documentclass[11pt,titlepage]{article}
\usepackage{graphics}
\usepackage{graphicx}
\usepackage{natbib}
\usepackage[bookmarks=false]{hyperref}

\newcommand{\mshemail}{email: \texttt{handcock@stat.washington.edu}}

\def\IE{\hbox{I\kern-.1667emE}}

\hypersetup{
pdfauthor = {Mark S. Handcock},
pdftitle = {On Sexual contacts and epidemic thresholds},
pdfsubject = {statistical inference for networks},
pdfkeywords = {statistical inference, networks, epidemiology, degree distribution, Yule, Waring},
pdfcreator = {Mark S. Handcock's brew},
pdfproducer = {Mark S. Handcock's brew}
}

\begin{document}

\title{On ``Sexual contacts and epidemic thresholds,"
models and inference for Sexual partnership distributions}

\author{Mark S. Handcock\thanks{Corresponding Author:
Center for Statistics and the Social Sciences,
Box 354320,
Seattle, WA 98195-4320,
\mshemail,
Phone: 206-221-6874,
Fax: 206-221-6873} \\
Departments of Statistics and Sociology \\
Center for Statistics and the Social Sciences \\
Department of Statistics \\
University of Washington, Seattle, WA
 \and 
James Holland Jones \\
Center for Studies in Demography and Ecology \\
Center for Statistics and the Social Sciences \\
Center for AIDS and Sexually Transmitted Diseases \\
University of Washington, Seattle, WA
 \and 
Martina Morris \\
Departments of Sociology and Statistics \\
Center for Studies in Demography and Ecology \\
University of Washington, Seattle, WA
}

\date{27 May 2003\\
Working Paper \#31 \\
Center for Statistics and the Social Sciences \\
University of Washington \\
Box 354322 \\
Seattle, WA 98195-4322, USA}

\maketitle

\begin{abstract}

Recent work has focused attention on statistical inference for
the population distribution of the number of sexual partners 
based on survey data.

The characteristics of these distributions are of interest as
components of mathematical models for the transmission dynamics of
sexually-transmitted diseases (STDs). Such information can be used
both to calibrate theoretical models, to make predictions for real
populations, and as a tool for guiding public health policy.

Our previous work on this subject has developed likelihood-based
statistical methods for inference that allow for low-dimensional,
semi-parametric models.  Inference has been based on several proposed
stochastic process models for the formation of sexual partnership
networks. We have also developed model selection criteria to choose
between competing models, and assessed the fit of different models to
three populations: Uganda, Sweden, and the USA.
Throughout this work, we have emphasized the correct assessment of
the uncertainty of the estimates based on the data analyzed. We have
also widened the question of interest to the limitations of inferences
from such data, and the utility of degree-based epidemiological models
more generally.

In this paper we address further statistical issues that are important
in this area, and a number of confusions that have arisen in
interpreting our work. In particular, we consider the use of cumulative lifetime
partner distributions, heaping and other issues raised by Liljeros {\it et al.}
in a recent working paper.

For the full paper in PDF click on
\href{http://www.csss.washington.edu/Papers/wp31.pdf}{http://www.csss.washington.edu/Papers/wp31.pdf}

\end{abstract}

\end{document}